\documentclass{article}
\usepackage{graphicx}
\title{Volume Weighting in the No Boundary Proposal  }
\author{ S. W. Hawking}
\date{July 2007}
\begin{document}
   \maketitle

\begin{abstract} 
It has been suggested that the no boundary proposal would predict little or no slow roll inflation leading to an empty deSitter
universe. However it is argued that the probability  for the whole universe should be multiplied by a zero mode factor e$^{3N}$ which count the the number of Hubble volumes in the universe. This voice weighting is similar to to that in eternal inflation but derived a gauge invariant manner. It predicts that inflation began at a saddle point in the potential and that the universe was always in the semi-classical regime.  
\end{abstract}
\bigskip

Cosmology has no predictive power   without a theory of initial conditions. Because of the singularity theorems of Penrose and myself, many people assume that the initial state is necessarily of trans-Planckian curvature.  We have no ideas of how to formulate initial conditions in such situations.  String theory, at least in the form we know it, is based on perturbations about flat space, and so would break down along with classical general relativity. However, the singularity theorem relevant to cosmology, though not that for black holes, depends on the strong energy condition: 
\[T_{ab} V^a V^b \geq \frac{1}{2} T V^a V_a \]
for any timelike or null vector $V^a$.  This is always satisfied by gauge fields, but can be violated locally by scalar fields. It is therefore possible for the universe either to bounce or to approach a de Sitter state in the past. Such non-singular solutions form only a small subset of the space of all scalar field gravity solutions, but I shall show that the no boundary condition implies that they provide the dominant contribution to the present state. The curvature of the universe  need  never have been at the Planck level, and the birth of the universe can have been  entirely within the semi classical domain. String theory is not necessary for cosmology. 

The no boundary condition gives a measure on solutions for the universe, which seems to be heavily biased towards little or no inflation. However, I shall argue that the true measure of the universe is given by the amplitude, times a volume factor. This volume weighting restores the probability of high inflation, starting at a saddle point of the potential. 

Our best guess for the structure of spacetime at the present time, is that it is approximately of the form, 
\[M = X \otimes Y \]
where $X$ is four dimensional Minkowski space, and $Y$ is a six or seven dimensional internal space. The geometry of the internal space will determine the effective particle physics theory at low energy.  The metric of $Y$ will be Ricci flat at tree level, and will depend on a finite number of parameters, or moduli. However, one would expect quantum corrections and super symmetry breaking to remove the degeneracy, and introduce an effective potential, $V$, which was a function on the moduli space of $Y$. If $\phi$ are local coordinates on $Y$, they can be regarded as scalar fields on $X$. The potential, $V$, could have a large number of local minima, corresponding to a landscape of possible vacuum states of M theory.  So what is it that determines that we are in the standard model state, and not one of the possible alternative vacuum states?  

\begin{figure}[!hb]
\begin{center}
\includegraphics[width=11cm, trim=1mm 1mm 1mm 1mm, clip]{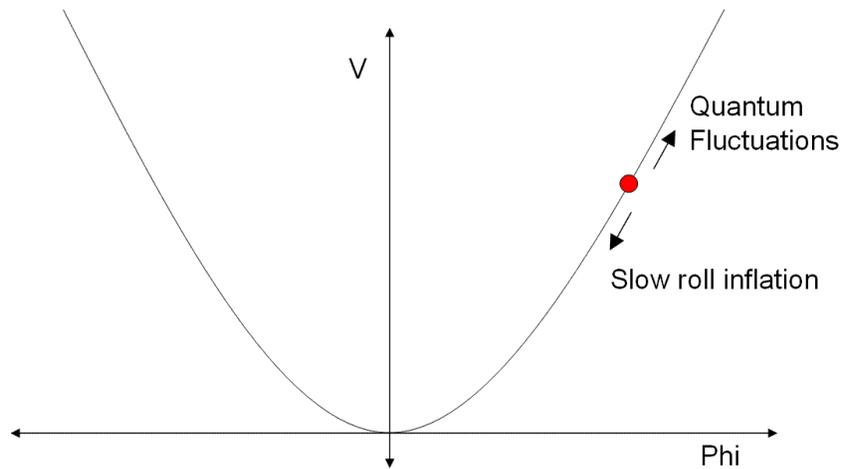}
\\
\caption{The Eternal Inflation Scenario: widely separated regions were supposed to fall into different local minima of the potential which would give the universe a mosaic structure.\label{the_only_one}}
\end{center}
\end{figure}

To answer this, one has to turn to cosmology.  One idea that has been advanced is the Eternal Inflation Scenario.  In this, the scalar fields, $\phi$, are supposed to fluctuate up in some regions, and down in others (see figure~\ref{the_only_one}). There will be as many regions in which it fluctuates down the potential hill as there are in which it fluctuates up, but the regions that fluctuate up, will expand faster. The upwards fluctuating regions will dominate later surfaces of constant time  if a certain condition is met: the condition for eternal inflation.  Widely separated regions that fluctuate  down the potential hill would fall into different local minima of the potential, which would give the universe a mosaic structure, with different parts of the universe in different vacuum states. This derivation of eternal inflation is not gauge invariant, and violates energy conservation and the Hamiltonian constraint.  However I will derive a similar condition from a very different argument.

Eternal inflation is an essentially classical picture, which assumes there is a single metric for the universe. That is why its advocates feel it is necessary to suppose the universe has a mosaic structure, to accomodate the possibility that the universe could be in any of the vacuum states, in the same universe.    However in a fully quantum theory, the universe can have any metric with suitable boundary conditions, which I shall take to be the no boundary condition.
\[ \Psi [ h_{ij}, \phi ] = \int{D g \mbox{e}^{-S[g]}} \]
The amplitude for a state with metric $g$, and matter field's $\phi$, on a spacelike surface $S$, is given by the path integral over all no boundary metrics, with those values on the surface $S$. One can also calculate the amplitude for inhomogeneous final states which are a mosaic of different vacuum states.  They will in general be lower than the amplitudes for homogeneous final states.

The amplitude, $\Psi$, is the wave function of the universe.  It will obey the Wheeler DeWitt equation:  
\[ \left[ -G_{ijkl} \frac{\delta^2}{\delta h_{ij} \delta h_{kl}} - ^3{}R(h) h^{\frac{1}{2}} + 2 \Lambda h^{\frac{1}{2}} \right] \Psi[h_{ij}] = 0 \]
where $G_{ijkl}$ is the metric on superspace,
\[ G_{ijkl} = \frac{1}{2} h^{-\frac{1}{2}} ( h_{ik}h_{jl} + h_{il}h_{jk} + h_{ij}h_{kl} )   \]
and $^3{}R$ is the scalar curvature of the intrinsic geometry of the three-surface.

In the case that the surfaces, $S$, are three-spheres of radius $a$, and the matter is a single scalar field $\phi$, this is a wave equation in the $(a, \phi)$ plane, with $a$ playing the role of time. 

\[ \frac{1}{2} \left[ \frac{\partial^2}{\partial a^2} - a^2 - \frac{1}{a^2} \frac{\partial}{\partial \phi^2} + a^4 V \right] \Psi(a,\phi) = 0 \]

In the Euclidean region, $a^2{}V <1$, there will be a real Euclidean solution of the field equations, and the wave function  will be exponential. Outside this region, however, there will only be complex solutions, and the wave function  will oscillate rapidly. One can represent the wave function as the product of a rapidly varying phase, $C$, and slowly varying amplitude, $B$.  Plugging this in the Wheeler DeWitt equation, one finds that $C$ obeys the Hamilton Jacobi equation.
\begin{eqnarray*}
 \Psi = B \mbox{e}^{iC}						\\
 \bigtriangledown{C} \cdot \bigtriangledown{C}-J=0	\\
 \bigtriangledown{B} \cdot \bigtriangledown{C}=0
\end{eqnarray*}

 One can therefore interpret the wave function by WKB, as corresponding to a family of Lorentzian solutions of the field equations.  The trajectories of the solutions are given by the gradient of $C$, raised by the Wheeler DeWitt metric. The amplitude, $B$, obeys a conservation equation, which implies that the amplitudes of individual solutions are constant over the evolution of the solutions. 

\[ a = H^{-1} \cosh(Ht) \mbox{, where } H^2 = V_1 \mbox{.}\]

The wave function of the universe given by the no boundary proposal, corresponds to solutions that bounce at the boundary of the Euclidean region. The potential will be a maximum at the minimal surface. The amplitude of the solution will be the amplitude of a $K = 0$ surface, at the potential at the bounce. There will be a mismatch in the derivatives of the scalar field, but this will be small if the potential satisfies the slow roll condition, $\bigtriangledown V$ small compared to $V$. 

The amplitude of a solution will be e$^{_I}$, where $I = -\frac{3}{4} \pi V$, is the Euclidean action of half a four-sphere with curvature scalar, $R = 8 \pi V$, and $V$ is the value of the potential at the bounce. The amplitude will be a maximum for solutions that bounce at the minimum of the potential. However, such solutions will just be empty deSitter space, and so not candidates for the universe we observe. To obtain a matter filled universe with structure, like galaxies and stars, it seems necessary for the universe to have a large number, $N$, of $e$-foldings of slow roll inflation. If one weights solutions with their amplitude, the probability distribution would be strongly biased to low values of $N$, the number of $e$-foldings of inflation.  This would predict that  our universe would have  the least value of $N$ compatible with our existence, which would not produce the universe we observe. 

This has been recognized to be a problem with the no boundary proposal for some time. I think the answer, is that there are two different probabilities involved. The amplitude gives the probability for the entire universe. However, one does not observe the entire universe, but only a Hubble volume around oneself. The number of such Hubble volumes at a given matter density, is proportional to the volume of the universe at that time, which in turn is proportional to e$^{3N}$. Thus on a frequency definition of probability, the probability of observing a Hubble volume of a given history, is proportional to the probability of that history, times e$^{3N}$. 
\begin{eqnarray*}
 P(\mbox{Entire universe}) &= & |\Psi^2 |	\\
 P(\mbox{Hubble volume}) & = & |\Psi^2 | \mbox{e}^{3N}
\end{eqnarray*}

The volume weighting transforms the probability distribution over $N$, the amount of inflation. It can more than compensate for the reduction in amplitude, due to a higher value of the potential at the bounce, if the slow roll parameter, $\epsilon$, is less than the potential, $V$, in Planck units, 

\[ \epsilon = \frac{\bigtriangledown V \cdot \bigtriangledown V}{V^2} < V \]

This is the same as the condition for eternal inflation but derived in a gauge invariant manner. 

At the time the microwave fluctuations we observe left the horizon,  this condition is not because $\frac{V}{\epsilon}$ was about $10^{-5}$  then. But inflation may have started before that at higher $V$, and lower $\epsilon$. For solutions that start at  the Planck density, in a polynomial potential, this condition will be satisfied only near the Planck density. The probability distribution would still be overwhelmingly in favor of low $N$. On the other hand, for a solution that starts at a maximum or a saddle point, the probability distribution would favour very large $N$. 

The dominant contribution is likely to come from broad saddle points well below the Planck density. 
\[ \frac{V''}{V} > -2 \]
\[ \mbox{Amplitude} = \exp(V_1{}^{-\frac{1}{2}}) \]
The metrics will be well within the semi classical regime. They would start out with a Hawking Moss instanton, a de Sitter like state which is unstable, and begins to run down the potential hill.  The origin of the universe, is in the low energy regime of M theory, in which four dimensional general relativity is a good approximation. This is supported by the fact that calculations based on four dimensional general relativity, are in excellent agreement with observations of the microwave background. One would not expect this, if the four dimensional approximation, $X \otimes Y$, broke down before one gets back to the time of inflation. This would indicate that the internal space, $Y$, is smaller than $10^6$ times the Planck length. The only situation in which 4D general relativity would break down, and in which one would need string theory, or some other approach, would be the final stages of evaporation of a black hole.

The only vacuum states that will have significant amplitudes to be matter filled, will be those where the minimum of the potential, lies on the line of descent from a broad saddle point of index one. By this I mean that there is only one direction in which the action decreases. This is in accord with the general principle, that the instanton that describes the decay of an unstable state, should have one, and only one, negative mode. In this case, the instanton would be the Hawking Moss or de Sitter instanton, with the value of the vacuum energy at the saddle point. The negative mode, would be the homogeneous mode in which the scalar fields everywhere move along the line of steepest descent from the saddle point. If the saddle point is broad, that is, if $\frac{V''}{V}$, is small and negative, the lowest inhomogeneous mode will be positive. This will give the Hawking Moss instanton one, and only one, negative mode, as required.  

In the usual, bottom up, approach, one assumes that the universe started in a state of high symmetry, which then evolved to the present broken symmetry state. The symmetry breaking would happen in different directions in different places, leading to topological defects, such as domain walls, cosmic strings, and monopoles. On the other hand, according to the top down approach I have described, the solution that gives the dominant contribution to the amplitude, could have the same broken symmetry all the way back. In this case, there would be no production of topological defects.

The no boundary condition, enables us in principle to calculate volume weighted quantum amplitudes, for the whole universe to be in different states in the landscape. It would be a mistake to assume that we should be in the state with the highest volume weighted amplitude. That would be like saying I should be Chinese, because there are many more Chinese than Brits. If the volume weighted amplitude for the standard model vacuum is non zero, it is irrelevant what the volume weighted amplitudes for other vacuum states are. The theory can not predict a unique vacuum state.  Instead, we have to input that we live in the standard model vacuum.

The bouncing universes that the no boundary proposal predicts, might seem at first sight, similar to the Ekpyrotic or cyclic universes. However, there is an important difference.In the Ekpyrotic universe, it is implicitly assumed that the state in the infinite past, is one of minimum excitation, although this is not clearly stated, or well defined. This means perturbations would be small in the infinite past, and grow during collapse, and subsequent expansion. In other words, the thermodynamic arrow of time, will point forward in both the contraction, and the expansion. By contrast, the no boundary solutions will have minimum excitations at the bounce, where the no boundary condition is imposed. This means the arrow of time will point forward in the expanding phase, and backward in the contracting phase. 

In fact, the physical relevance of the contracting phase, is questionable. It is like the analytical continuation of the semi-classical solution that describes pair creation in an electric field.. This consists of an electron and a positron that come in from infinity at $t < 0$, are brought to rest at $t = 0$, and accelerate away from each other for $t > 0$. However, physically there are no incoming particles.  Instead one says that the electron positron pair was quantum created at $t = 0$. In a similar way, one should not attach any physical significance to the early contracting phase of the universe, but say the universe was quantum created at the bounce. 


The no boundary proposal, and the top down approach, allow us to calculate amplitudes for different states at the present time. These overwhelmingly favor low or zero inflation, which would lead to an almost empty deSitter like universe. However, I argue that the probability of the entire universe, given by the amplitude, should be multiplied by the volume, to get the probability of observing a Hubble volume. This volume weighting favors inflation starting at a saddle point of the potential, and leads to the prediction  that the universe should be flat, within the limits of the fluctuations in the microwave background. 
 
The dominant histories in the path integral for these amplitudes, are bouncing solutions of the field equations, which lie entirely in the semi classical regime of four dimensional general relativity. However, these should probably not be interpreted as describing bouncing universes, but rather as the quantum creation of universes in de-Sitter like states. 

The amplitudes will be highest for states in which the whole universe is in a single state, rather than a mosaic of different states, as predicted by eternal inflation. There will be no primordial production of topological defects, such as monopoles, and cosmic strings. Not all states in the landscape will have significant amplitudes, but there will be more than one that do, so M theory does not predict a unique low energy particle physics theory.  It is implausible that life is possible only in one of these states, so we might have chosen a better location.


\end{document}